\documentstyle[preprint,tighten,aps,prd]{revtex}
\begin{document}

\title
{Do Quarks Really Form Diquark Clusters in the Nucleon?}

\author{Derek B. Leinweber}

\address{
Department of Physics and Center for Theoretical Physics \\
University of Maryland, College Park, MD 20742
}

\maketitle

\begin{abstract}
A gauge invariant method for the investigation of scalar diquark
clustering in the nucleon ground state is presented.  The method
focuses on a comparison of quark distributions in the nucleon with
those in the $\Delta$ baryon resonance.  Recent lattice QCD
calculations of these quark distribution radii are analyzed in a
search for evidence of scalar diquark clustering.  The analysis
indicates the lattice results describe the negative squared charge
radius of the neutron with little resort to hyperfine clustering
between $u$-$d$-quark pairs.  This result contrasts both quark-diquark
and nonrelativistic quark models where hyperfine attraction between
$u$ and $d$ quarks in the nucleon is argued to play a significant
role.  Comparison of light quark distributions in $\Lambda^0$ and
$\Sigma^{*0}$ indicate only a small reduction of the scalar diquark
distribution radius relative to the vector diquark distribution.
Current lattice QCD determinations of baryon charge distributions do
not support the concept of substantial $u$-$d$ scalar diquark
clustering as an appropriate description of the internal structure of
the nucleon.
\end{abstract}


\narrowtext

\section{INTRODUCTION}

   The one-gluon-exchange potential (OGEP) has been extensively used
to describe the spin-dependent interactions of constituent quarks in
low-energy phenomenology since its inception \cite{derujula75}.  Among
the earliest of OGEP successes is an explanation of the negative
squared charge radius of the neutron.  Carlitz {\it et al.} focused on
the spin-dependent hyperfine repulsion between the doubly represented
$d$ quarks in the neutron which naturally gives rise to a negative
squared charge radius \cite{carlitz77}.  Later analyses exploited the
larger OGEP hyperfine attraction acting between constituent quark
pairs in a scalar spin-0 state \cite{isgur78,isgur81}.  In these
papers the attractive hyperfine force pulls the $u$-quark of the
neutron into the center while the two $d$-quarks are repelled to the
periphery by the vector spin-1 repulsion of the hyperfine interaction.
These more quantitative analyses were also successful in accounting
for the neutron charge radius.

   The success of hyperfine interactions based on single-gluon
exchange, led some authors to speculate that the strong attractive
hyperfine interaction pairs a $u$ and $d$ quark of the nucleon into
a scalar diquark \cite{lichtenberg69} of considerable clustering
\cite{sateesh92,anselmino91,efimov90,dziembowski90,dosch89,fredriksson84}.
It is often argued that point-like scalar diquarks naturally arise in
QCD, or that scalar diquarks are energetically favored
\cite{fredriksson83,fredriksson82}.  A factor in the popularity of
diquark models is the numerical simplification of three body problem
to a two body problem \cite{efimov90,lichtenberg82,lichtenberg83}.
Also the diquark model is not without a long list of successful
descriptions of low-energy hadronic phenomena.  For example, diquark
clustering can also explain the neutron charge radius.  In this
approach it is concluded that the neutron has a scalar diquark core
\cite{dziembowski81}.  More recently it has been argued that a
quark-diquark approximation of the three-quark structure of baryons is
now available which takes into account the inner structure of baryons
at low energies \cite{efimov90}.  Still others argue that diquark
models provide a natural forum for investigating deviations from
$SU(6)$-spin-flavor symmetry as low energy phenomena is not tractable
{}from first principles \cite{uppal91}.

   In this paper, a gauge invariant method for the examination of
scalar diquark clustering in the nucleon ground state is presented.
Direct evidence indicating the absence of substantial diquark
clustering of $u$ and $d$ quarks in low-lying baryons will be
presented.  Quark electric charge distribution radii are calculated
{}from first principles in lattice regularized QCD.  By examining the
quark distributions in octet baryons where quarks may form scalar
diquarks and comparing these distributions with the relevant decuplet
baryons where quarks are predominantly paired in vector diquarks, one
can determine the amount, if any, of scalar diquark clustering
\cite{gaugeinv}.  A central point of this paper is that it is a
comparison of charge distribution radii between octet and decuplet
baryons (as opposed to within the baryon octet) that reveals whether
or not hyperfine interactions lead to scalar diquark clustering.

   Of course this is not the first paper to refute the diquark picture
of low-energy baryons.  As early as 1981, an analysis of $\pi N$
partial widths suggested that diquark configurations do not contribute
appreciably to the structure of low-lying resonances \cite{forsyth81}.
More recently it has been argued that point-like scalar diquarks are
unlikely in a relativistic bound-state quark model \cite{ram87} and
that there is no clear indication of diquark clustering in a
non-relativistic quark model of the nucleon \cite{fleck88}.

   In some systems there is good reason to believe that some diquark
clustering can occur.  For example, in $\Xi$ the two strange quarks
are more localized due to their heavier mass \cite{leinweber91}.  In
high angular momentum systems, small attraction between quarks can
lead to clustering \cite{fleck88}.  However, the scalar diquark
clustering argued to arise from the attractive part of the hyperfine
interaction is not reproduced in the nonperturbative analysis
presented here.

   In section II A, descriptions of the neutron charge radius in a
quark-diquark model, nonrelativistic constituent quark model and
lattice QCD calculations are compared.  The purpose of this section is
to illustrate that the three different approaches lead to values for
the neutron to proton charge radius ratio which are consistent with
each other and that the lattice results are not inconsistent with the
experimentally determined value.  The conclusion of this section is
that reproduction of the neutron to proton charge radius ratio alone
is insufficient to discriminate between the various quark distribution
pictures.  Section II B introduces the charge distributions in
$\Delta$ to discriminate between these three different descriptions of
quark distributions in the nucleon.  Here the absence of significant
scalar diquark clustering in the lattice results is discussed.
Section III explores charge distributions in $\Lambda^0$ and
$\Sigma^{*0}$ hyperons.  A similar analysis indicates the absence of
substantial scalar diquark clustering in these hyperons.  Section IV
considers the possible sources of systematic uncertainty in the
lattice calculations and how these uncertainties may affect the charge
distribution radii used in this analysis.  Finally the implications of
the results are discussed and summarized in section V.

\section{SCALAR DIQUARK CLUSTERING IN THE NUCLEON}

\subsection{The Neutron Charge Radius}

   To investigate the issue of scalar diquark clustering, we turn to
the lattice investigations of baryon electromagnetic structure of
Ref. \cite{leinweber91,leinweber92b}.  These lattice results are
obtained in a numerical simulation of quenched QCD on a $24 \times 12
\times 12 \times 24$ lattice at $\beta=5.9$ using Wilson fermions.
Twenty-eight gauge configurations are used in the analysis.  Charge
radii are obtained from lattice calculations of electromagnetic form
factors at $\vec q\,^2 = 0.16$ GeV${}^2$. The $Q^2$ dependence of the
form factors is taken as a dipole form.  However, monopole or linear
forms yield similar radii which agree with the dipole results within
statistical uncertainties.  A third order, single elimination
jackknife \cite{efron79,gottlieb86} is used to determine the
statistical uncertainties in the lattice results.  These uncertainties
are indicated in parentheses describing the uncertainty in the last
digit(s) of the results.

   The lattice results have displayed some of the qualitative features
expected from hyperfine interactions.  Let us first focus on the mean
square charge radius of the neutron as described by the quark-diquark
model \cite{dziembowski81}, the nonrelativistic quark model
\cite{isgur78,isgur81} and the lattice simulations of Ref.
\cite{leinweber91,leinweber92b}.  The diquark description of quark
distributions is as illustrated in figure 1a.  A $u$ and $d$ quark
form a scalar diquark cluster whose radius in extreme cases may be
taken to be point like.  The second $d$ quark has a larger charge
distribution radius and provides the required negative charge at large
radii surrounding a positive core of charge +1/3.

   In the nonrelativistic quark model a similar scalar diquark
clustering occurs between $u$ and $d$ quarks due to the attractive
part of the hyperfine interaction.  However this clustering is
relatively small compared to the clustering anticipated in diquark
models.  The remaining $d$-quark is driven to the periphery of the
neutron by the repulsive part of the hyperfine interaction.  The
clustering is averaged over the two $d$ quarks of the neutron and the
quark distributions may be described as in figure 1b.  An important
point to note is that the hyperfine interaction causes {\it two}
alterations to the unperturbed wave function.  The attraction between
the $u$ and $d$ quarks causes the whole nucleon to shrink and the
repulsion between the two $d$ quarks introduces the required asymmetry
between the quark distributions to reproduce the neutron charge
radius.

   In understanding the results of the lattice investigation the
negative mean square charge radius of the neutron was attributed to
hyperfine repulsion between the doubly represented $d$ quarks without
resorting to any hyperfine attraction \cite{leinweber91,draper90}.
This argument is similar to the original explanation provided by
Carlitz {\it et al.} \cite{carlitz77}.  Of course it is the hyperfine
{\it attraction} that is argued to naturally give rise to scalar
diquark clustering in the nucleon.  The current picture of quark
distributions within the neutron is described as in figure 1c.  The
difference between figures 1b and 1c is the possible absence of any
hyperfine attraction between $u$ and $d$ quarks.  Of course the
lattice results for the neutron charge radius alone may also be
consistent with either of the previous two descriptions
\cite{insuffinfo}.

   The presence of two degrees of freedom, namely the matter radius
and the charge asymmetry, allow these three different descriptions of
quark distributions within the neutron to lie in quantitative
agreement when reproducing the neutron to proton charge radius ratio.
Experimental measurements \cite{hohler76,dumbrajs83} of nucleon mean
square charge radii produce a ratio of
\begin{equation}
{\bigm < r^2 \bigm >_n \over \bigm < r^2 \bigm >_p} = -0.167(7) \, .
\end{equation}
The quark-diquark prediction of the mean square neutron to proton
charge radius ratio is
\begin{equation}
{\bigm < r^2 \bigm >_n \over \bigm < r^2 \bigm >_p} = -0.137(9) \, ,
\end{equation}
when the parameter of the model is fixed by previous analyses
\cite{dziembowski81}.  Similarly the configuration mixing induced by
hyperfine interactions in the nonrelativistic quark model
\cite{isgur78} leads to the result
\begin{equation}
{\bigm < r^2 \bigm >_n \over \bigm < r^2 \bigm >_p} = -0.13 \, .
\end{equation}

   Present lattice simulations calculate hadron properties at
relatively heavy current quark masses due to the increasing
computational demands encountered in inverting the fermion matrix as
the quarks become lighter.  In Ref. \cite{leinweber91,leinweber92b}
charge radii are determined at three values of the Wilson hopping
parameter $\kappa = 0.152$, 0.154, and 0.156 corresponding to current
quark masses ranging from the strange current quark mass $m_s$ through
to $m_s/2$.  To make contact with the physical world the lattice
results are extrapolated, usually linearly as a function of $1/\kappa$
which is proportional to the current quark mass or $m_\pi^2$, to the
point at which the physical pion mass is reproduced.  Results from
chiral perturbation theory \cite{beg72,leinweber92c,cohen92a} suggest
that logarithmic terms divergent in the limit $m_\pi \to 0$ should be
included in the analytic structure of the nucleon charge radius
extrapolation function, in addition to terms linear in $m_\pi^2$ and
other higher order terms finite in the chiral limit.  However, the
physics of light pions giving rise to the divergent logarithmic term
is not included in either of the quark models considered here.  To
allow a comparison with these models on a more equal footing, the
lattice extrapolations of charge radii are done linearly in $m_\pi^2$,
effectively subtracting the nonanalytic contributions associated with
light pion dressings from the charge radii.  For linear
extrapolations, the difference between extrapolations of quantities to
the physical pion mass as opposed to $m_\pi = 0$ are negligible.  In
the following, radii are extrapolated to $\kappa_{\rm cr} =
0.159\,8(2)$ where the pion mass vanishes.

   The neutron charge radius is sensitive to gluonic degrees of
freedom and the lattice ratio has a large statistical error associated
with it.  However, it is important to demonstrate that the lattice
result is consistent with the previous analyses.  The lattice results
indicate \cite{leinweber92b}
\begin{equation}
{\bigm < r^2 \bigm >_n \over \bigm < r^2 \bigm >_p} =
       -0.11 \left ( {\textstyle{+0.10 \atop -0.14}} \right ) \, .
\end{equation}

   The qualitative features of the neutron charge radius anticipated
by hyperfine interactions are reproduced in each of the calculations
considered.  Obviously, reproduction of the neutron to proton charge
radius ratio alone is not sufficient to determine whether it is
hyperfine attraction, repulsion, or a combination of the two that is
actually responsible for giving rise to the neutron charge radius.  If
attraction plays a significant role then diquark-clustering models may
capture the essential features of the underlying quark-gluon dynamics.
On the other hand, if the lattice results indicate hyperfine
attraction does not play a significant role in determining the charge
distributions then the realization of diquark clustering in the
nucleon ground state is doubtful.

\subsection{Charge Distributions in $\Delta$}

   To discriminate between these three different descriptions of quark
distributions within the nucleon we must turn to another system where
the attractive part of the hyperfine interaction does not play a
significant role, {\it i.e.} the $\Delta$ baryon resonance.  By
examining the changes in the quark distributions as the spin of the
singly represented quark is set symmetric with the doubly represented
quarks we can investigate the relevance of scalar diquark clustering
and search for the anticipated effects of one-gluon-exchange hyperfine
interactions on the quark distributions.

   Consider the quark distributions within the proton and how they
will change in going from $p$ to $\Delta^+$.  Figure 2 illustrates the
three previous scenarios for the proton as well as the quark
distributions in $\Delta^+$.  Hyperfine interactions are not expected
to give rise to vector diquark clustering.  Both the nonrelativistic
quark model and the lattice results describe the $\Delta^+$ charge
distribution as a sum of three equivalent quark distributions.  In
fact the symmetry of the $\Delta$ three-point correlation function
demands the quark distribution radii in $\Delta$ to be equal under
$SU(2)$-isospin symmetry.  This symmetry is manifest without resorting
to actual calculations on the lattice \cite{leinweber92b}.

   In the quark-diquark model the scalar diquark cluster is lost in
$\Delta^+$ and the net positive charge of the cluster in the nucleon
moves to larger radii.  For point-like diquarks the net effect is
huge, resulting in a much larger charge radius for $\Delta^+$ than for
the proton.  Moreover, both $u$- and $d$-quark charge distributions
swell in $\Delta$ due to the breakup of the (point-like) $u$-$d$-quark
cluster.  Unfortunately quantification of these statements does not
appear to be possible.  A comparison of $p$ and $\Delta^+$ charge
radii in a diquark model does not appear to have been considered.
Some form factors of octet and decuplet baryons were recently examined
in the quark-diquark approximation \cite{efimov90}, however electric
form factors and charge radii for $\Delta$ resonances were not
reported.  To estimate a lower bound for the size of these anticipated
quark distribution swellings, we refer to the nonrelativistic quark
model where the role of hyperfine attraction plays a much weaker and
less dramatic role.

   In the nonrelativistic quark model, both the $u$- and $d$-quark
distributions become broader as the attraction between $u$ and $d$
quarks is replaced by hyperfine repulsion in $\Delta$.  In the model
of Isgur-Karl-Koniuk \cite{isgur78} the predicted increase in the rms
charge radius is \cite{isgur80,qmnote}
\begin{equation}
{r_\Delta \over r_p} = 1.28 \, ,
\end{equation}
with the quark distributions experiencing a large swelling of
\begin{equation}
{r_\Delta^u \over r_p^u} = 1.33 \, , \quad
{r_\Delta^d \over r_p^d} = 1.49 \, .
\end{equation}

   In the scenario previously introduced for the lattice results, the
dominant effect will be the new hyperfine repulsion experienced by the
$d$ quark from the two $u$ quarks in $\Delta^+$.  Similarly the two
$u$ quarks already experiencing some hyperfine repulsion will feel
additional repulsion from the single $d$ quark.  Hence the
dominant effect will be the broadening of the negatively charged
$d$-quark distribution compensated by some broadening of the $u$-quark
distribution.  In fact, the lattice results suggest that the $\Delta^+$
charge radius may actually be {\it smaller} than that of the proton
\begin{equation}
{r_\Delta \over r_p} = 0.97(7) \, .
\end{equation}
This result differs by 4 standard deviations from the prediction of
the nonrelativistic quark model.  The quark distribution radii
indicate the dominant effect in the lattice results is the broadening
of the negatively charged $d$-quark distribution
\begin{equation}
{r_\Delta^u \over r_p^u} = 1.01(9) \, , \quad
{r_\Delta^d \over r_p^d} = 1.12(16) \, .
\label{pdelqradrat}
\end{equation}

   The striking difference between the lattice results and the two
models considered here is {\it the absence of any significant change
in the lattice $u$-quark distribution radius}.  In both models, the
$u$-quark distribution was predicted to be broader in $\Delta$,
largely due to the disappearance of scalar diquark clustering in going
{}from the nucleon to $\Delta$.  The lattice results indicate that
hyperfine attraction does not lead to substantial scalar diquark
clustering in the nucleon ground state \cite{attrvsrepul}.

   Figure \ref{pdelrad} illustrates the extrapolation of quark
distribution radii in $p$ and $\Delta^+$ used to obtain the results of
(\ref{pdelqradrat}).  The $u$-quark distribution radii in $p$ and
$\Delta$ are nearly identical for each $\kappa$ considered.  Table
\ref{protdelta} details the charge distribution radii for $p$ and
$\Delta$ and their residing quarks.

\section{SCALAR DIQUARK CLUSTERING IN $\Lambda^0$}

   Another place to search for evidence of scalar diquark clustering
is in the charge distribution radius of the light $u$ and $d$ quarks
in $\Lambda^0$.  In simple quark models these quarks are generally
taken as a pure scalar diquark.  In other words, in this approximation
the $\Lambda^0$ magnetic moment is given by the intrinsic magnetic
moment of the strange quark alone.  The lattice results suggest that
the $u$ and $d$ quarks do not form a {\it pure} scalar diquark and may
actually contribute to the $\Lambda^0$ magnetic moment at the level of
10\%.  However, for the present purpose we will ignore such effects.

   In the decuplet $\Sigma^{*0}$ hyperon state the $u$-$d$ sector will
be paired predominately as a vector diquark.  Hence with the
assumption that the $s$ quark plays a spectator role, comparison of
the light quark sector distribution radii in $\Lambda^0$ and
$\Sigma^{*0}$ will give some indication of the relevance of scalar
diquark clustering.  If a scalar diquark is ``dissolved'' in going
{}from $\Lambda^0$ to $\Sigma^{*0}$, a broadening of the light quark
distribution in $\Sigma^{*0}$ is expected.  The ratio of the light
quark distributions from the lattice analyses
\cite{leinweber91,leinweber92b} indicates
\begin{equation}
{r_{\Sigma^{*0}}^l \over r_\Lambda^l} = 1.04(10) \, ,
\label{lamsigqradrat}
\end{equation}
confirming the conclusions based on the N-$\Delta$ quark
distributions.  Once again substantial diquark clustering is not
seen.

   Figure \ref{lamsigrad} illustrates the extrapolation of quark
distribution radii in $\Lambda$ and $\Sigma^{*0}$ used to obtain the
results of (\ref{lamsigqradrat}).  The strange quark appears to play a
satisfactory spectator role.  Little change is seen between the
combined light-quark distributions of $\Lambda$ and $\Sigma^{*0}$ at
each value of $\kappa$.  Numerical values and statistical
uncertainties for these radii are summarized in table
\ref{lambdasigma}.

\section{SYSTEMATIC UNCERTAINTIES}

   This comparison of charge distributions in octet and decuplet
baryons has revealed remarkable differences between the lattice
results and the anticipated role of hyperfine attraction based on the
OGEP.  Moreover, this analysis appears to render a diquark picture of
quark distributions in the nucleon ground state obsolete.  As a
result, it is important to consider the possible sources of systematic
uncertainty in the lattice calculations.  Systematic uncertainties may
have their origin in the quenched approximation, the extrapolation of
the lattice results to the physical regime, finite volume effects or
renormalization group scaling deviations.

   It is worth noting that all of the lattice results presented to
this point have involved ratios of lattice results.  It is expected
that the effects of these possible sources of systematic uncertainty
will be suppressed in taking ratios.  For example, while the absolute
values of the lattice predictions of magnetic moments
\cite{leinweber91} are small compared to the experimentally measured
values, the lattice results predict ratios of the baryon magnetic
moments to the proton moment as good as or better than models, which
in most cases have parameters tuned to reproduce the experimental
moments.  A more detailed discussion of systematic uncertainties
follows.

   A calculation of electromagnetic form factors in full QCD has not
been done.  However, at the present values of quark mass investigated
on the lattice, hadron spectrum analyses suggest the dominant new
physics in full QCD is a simple renormalization of the strong coupling
constant.  It is possible that over large distances the quenched
approximation does not screen quark interactions as much as required
by full QCD.  However given the similarity of the quark distribution
radii discussed in this paper it is unlikely that there would be
significant deviations from the results presented here.

   The extrapolation of the results to the physical regime is another
source of concern.  However, the arguments based on the extrapolated
results are supported at each value of quark mass investigated on the
lattice as indicated in tables \ref{protdelta} and \ref{lambdasigma}.
Because the actual lattice calculations are done with $u$- and
$d$-current-quark masses the order of the strange-current-quark mass,
one might be concerned that hyperfine interactions such as those of
the OGEP are suppressed in the lattice calculations to the point that
one has no hope of observing diquark clustering.  Fortunately, this is
unlikely to be the case.

   To assess this issue more quantitatively, one can resort to the
one-gluon-exchange hyperfine interaction term which is inversely
proportional to the product of quark masses.  An important point is
the OGEP has some relevance only in phenomenology where constituent
quark masses are used in the hyperfine term.  While the current quark
masses used in this investigation are somewhat heavy, the
corresponding constituent quark masses are not too different from that
used in phenomenology.  Perhaps the easiest way to quantify this is to
estimate the constituent quark mass to be one third the lattice proton
mass.  At the lightest quark mass considered on the lattice, the
constituent quark mass is approximately 430 MeV which is not too
different from the value obtained from the physical mass at 313 MeV.
If the hyperfine term of the one-gluon-exchange interaction is
relevant, then the differences in the strength of the hyperfine
interactions for these two quark masses are within a factor of 2.  For
the intermediate quark mass value the strength is within a factor of
2.5.  Hence, there is good reason to believe an extrapolation to the
physical value is adequate.  Moreover, significant mass splitting is
observed between the nucleon and $\Delta$ on the lattice where
$M_N/M_\Delta = 0.87(8)$.  However, the change in the $u$-quark
distribution radius remains negligible.  Perhaps it is also worth
noting that single-gluon-exchange interaction strengths between
current quarks in this lattice QCD calculation exceed the strengths
common to phenomenological analyses by an order of magnitude for the
lightest quark mass investigated on the lattice.

   It has been argued in the past that the finite volume of the
lattice may affect the lattice results, as surrounding periodic images
of the baryon under study may restrict the baryon's size
\cite{leinweber91}.  Note however, the $u$-quark distribution in
$\Delta$ is not the broadest quark distribution seen on the lattice.
In addition, similar effects are seen for heavier quark masses where
the quarks are more localized and less sensitive to the boundaries of
the lattice.  For a given value of $\kappa$, the charge distribution
radii calculated on the lattice tend to be similar in size.  For this
reason the finite volume of the lattice is more likely to affect the
lattice results in a global manner.  For the case of a scalar $u$-$d$
diquark cluster in $p$ breaking up in $\Delta^+$, a finite volume
effect could not simultaneously accommodate a swelling of the
$d$-quark distribution and yet restrict the $u$-quark distribution
{}from following a similar swelling.

   The issue of deviations from asymptotic scaling is difficult to
assess quantitatively since a similar calculation at finer lattice
spacings has not been done.  Toussaint \cite{toussaint92} has
demonstrated that the nucleon to $\rho$ mass ratio calculated at
$m_\pi/m_\rho = 0.6$ is, to a good approximation, independent of the
value of $\beta$ over a range of $\beta = 5.7$ to 6.3 for quenched
Wilson fermion calculations.  Furthermore, Yoshie {\it et al.}
\cite{yoshie91} have demonstrated that it is possible to reproduce the
hadron spectrum in quenched QCD at $\beta = 6.0$ within statistical
uncertainties of approximately 10\% \cite{toussaint92}.

   The lattice prediction of the current investigation for the
$\Delta/N$ mass ratio lies 12\% or 1.4 $\sigma$ below the experimental
ratio.  One could adjust the parameters of the nonrelativistic quark
model to reproduce the lattice $N$-$\Delta$ splitting in which case
the difference between the lattice and model predictions for baryon
charge radii is reduced to 2$\sigma$.  However, the trend of the
lattice results indicates that a doubling of the effects seen in the
lattice results only widens the gap between the predictions of
$\Delta^+/p$ charge radii ratios and leaves the change in the
$u$-quark radius in going from $p$ to $\Delta^+$ at the order of 2\%.

\section{DISCUSSION AND SUMMARY}

   Faced with the discrepancies of these models which both lack
mesonic degrees of freedom it is interesting to consider models such
as the cloudy bag \cite{thomas84}, or hedgehog models such as the
Skyrmion \cite{zahed86}, hybrid (or little) bag \cite{vepstas90} and
the chiral-quark meson model \cite{banerjee87}.  In hedgehog models,
the proton and $\Delta^+$ radii are equal by construction.  The
difference in charge radii are $1/N_c$ suppressed and cannot be
calculated using conventional semi-classical techniques.  It may be
useful to note that the lattice results suggest that such higher order
effects may be small.

   A calculation of the charge radii considered here in the cloudy bag
model could provide further insight to the importance of mesonic
degrees of freedom in describing baryon charge distributions.  Since a
large part of the neutron charge radius has its origin in the pion
cloud of the cloudy bag model \cite{theberge80,myhrer82}, it may be
possible to circumvent the problems encountered here with hyperfine
interactions.  Of course, to allow a direct comparison with the
lattice results and to avoid problems associated with open channel
physics, the cloudy bag analysis should be done with heavier pions.
The $\Delta^+$ charge radius and internal quark distributions provide
new additional information on the spin-dependence of quark
interactions.  The traditional models examined here do not reproduce
the lattice results and a description of these phenomena remains an
open challenge to models of QCD.

   Some have argued the existence of scalar diquarks based on large
momentum transfer phenomenology.  An analysis of scalar and vector
diquarks at large momentum transfers in lattice QCD may provide useful
insight.  The results presented here may be of interest to those
investigating large $Q^2$ phenomena using nucleon ground state wave
functions as a model input representing the nonperturbative parts of
the calculation \cite{dziembowski90,meyer91}.

   Finally, a few comments on the absence of significant quark
clustering in the lattice results relative to that anticipated by
quark models are in order.  The first and most obvious comment to make
is that baryon charge distributions are sensitive to the long distance
nonperturbative aspects of QCD.  It should not be too surprising to
find that hyperfine interactions based on a single-gluon-exchange
become virtually irrelevant in the nonperturbative regime.  It is well
known that the one-gluon-exchange hyperfine interaction has some
relevance only when the quark masses are taken to be constituent quark
masses.  Experimental mass splittings indicate it is not the relevant
interaction between current quarks.  There is an infinite class of
multiple-gluon-exchange diagrams to consider beyond the exchange of a
single gluon.  Moreover, these diagrams are equally important since
the theory is nonperturbative.

   A second and possibly more interesting point concerns the spin of
the singly represented quark in the nucleon.  From the lattice QCD
analyses of baryon magnetic moments
\cite{leinweber91,leinweber92b,leinweber92a}, it has become clear that
the behavior of the singly represented quark in octet baryons is very
different from the predictions of $SU(6)$-spin-flavor symmetry.
Briefly stated the lattice results indicate that the proton moment is
better described as
\begin{equation}
\mu_p = {4 \over 3} \mu_u - {1 \over 6} \mu_d \, ,
\end{equation}
where in constituent quark model language, the $d$ quark has its net
spin opposite that of the $u$ quarks about half as much as suggested
by $SU(6)$.  In relation to scalar diquark clustering the probability
of finding $u$ and $d$ quarks paired in a spin-0 state has been
reduced.  Once again the lattice results suggest that scalar diquark
degrees of freedom do not provide an appropriate description of the
internal quark structure of low-lying baryons.

   A gauge invariant method for the examination of scalar diquark
clustering in the nucleon ground state has been presented.  Results
{}from lattice QCD describing the distributions of quarks in the baryon
octet and decuplet have been analyzed in a search for evidence of
scalar diquark clustering.  The results presented here contrast the
predictions of the nonrelativistic quark model which has a relatively
small diquark clustering compared to that demanded in quark-diquark
models.  The lattice results do not support the concept of substantial
diquark clustering as an appropriate description of the internal
structure of low-lying baryons.

\acknowledgements

   I wish to thank Zbigniew Dziembowski for stirring my interest in
this subject.  I also thank Wojciech Broniowski, Tom Cohen, Manoj
Banerjee, Nathan Isgur and Simon Capstick for a number of interesting
and helpful discussions.  Financial support from the U.S. Department
of Energy under grant DE-FG05-87ER-40322 is gratefully acknowledged.



\begin{thebibliography}{10}
\bibitem{gaugeinv}
Note that an individual quark flavor distribution radius is a gauge invariant
  quantity. The radius may be regarded as a baryon charge radius in which the
  remaining quark flavors have zero electric charge.
\bibitem{insuffinfo}
To this point we do not have sufficient information to exclude the possibility
  of a significant role for hyperfine attraction in the lattice results.
  However, it will become clear through the analysis of $\Delta$-baryons that
  hyperfine attraction does not play a significant role in determining the
  quark charge distribution radii.
\bibitem{qmnote}
Here we have taken $\theta_\Delta = - \theta_p$, and $\phi_\Delta = 0$.
  Equation (13) of Ref. \cite{isgur78} has been adjusted as described in the
  errata.
\bibitem{attrvsrepul}
In the preceding arguments, the focus in the lattice scenario has been on
  hyperfine repulsion. However, the lattice results are also consistent with a
  mixture of hyperfine attraction and repulsion or even exclusive hyperfine
  attraction provided the clustering is small. The important point to
  draw from this examination of charge radii is that there is no evidence of
  significant $u$-$d$-quark clustering in the nucleon ground
  state quark distributions.
\end{thebibliography}

\mediumtext
\begin{table}
\caption{Rms charge distribution radii normalized to unit charge in
         lattice units \break $\bigm < r^2/a^2 \bigm >^{1/2}$.}
\label{protdelta}
\setdec 0.00(00)
\begin{tabular}{lcccc}
Baryon &$\kappa_1=0.152$ &$\kappa_2=0.154$ &$\kappa_3=0.156$
       &$\kappa_{cr}=0.159\,8(2)$\\
\tableline
$p$            &\dec 3.70(13) &\dec 4.04(18)
               &\dec 4.39(33) &\dec 5.06(48) \\
$\Delta^{+} $  &\dec 3.71(13) &\dec 4.02(19)
               &\dec 4.34(39) &\dec 4.90(57) \\
$u_p$          &\dec 3.69(12) &\dec 4.00(18)
               &\dec 4.28(38) &\dec 4.86(57) \\
$u_{\Delta^+}$ &\dec 3.71(13) &\dec 4.02(19)
               &\dec 4.34(39) &\dec 4.90(57) \\
$d_p$          &\dec 3.63(14) &\dec 3.86(21)
               &\dec 3.99(66) &\dec 4.39(82) \\
$d_{\Delta^+}$ &\dec 3.71(13) &\dec 4.02(19)
               &\dec 4.34(39) &\dec 4.90(57) \\
\end{tabular}
\end{table}

\begin{table}
\caption{Rms charge distribution radii normalized to unit charge in
         lattice units \break $\bigm < r^2/a^2 \bigm >^{1/2}$.
$l_\Lambda$ indicates the combined light $u$ and $d$ quarks in
$\Lambda^0$ and similarly for $\Sigma^{*0}$.}
\label{lambdasigma}
\setdec 0.00(00)
\begin{tabular}{lcccc}
Baryon &$\kappa_1=0.152$ &$\kappa_2=0.154$ &$\kappa_3=0.156$
       &$\kappa_{cr}=0.159\,8(2)$\\
\tableline
$l_\Lambda$    &\dec 3.65(13)  &\dec 3.94(18)
               &\dec 4.18(44)  &\dec 4.67(73)  \\
$l_{\Sigma^*}$ &\dec 3.71(13)  &\dec 4.03(18)
               &\dec 4.40(33)  &\dec 4.99(48)  \\
$s_\Lambda$    &\dec 3.70(13)  &\dec 3.66(15)
               &\dec 3.59(22)  &\dec 3.51(29)  \\
$s_{\Sigma^*}$ &\dec 3.71(13)  &\dec 3.68(14)
               &\dec 3.65(20)  &\dec 3.60(28)  \\
\end{tabular}
\end{table}

\narrowtext

\begin{figure}
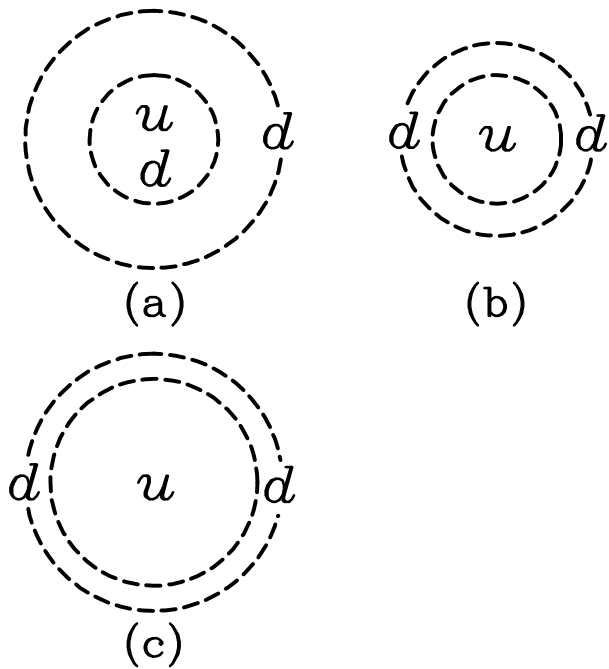

\caption{Sketches of quark distribution radii in the neutron as
described by $a)$ the quark-diquark model, $b)$ the nonrelativistic
quark model and $c)$ the lattice investigations of baryon
electromagnetic form factors.  The dashed lines are representative of
the rms charge radius of the quark distributions normalized to unit
charge.
\label{neutron}}
\end{figure}

\begin{figure}
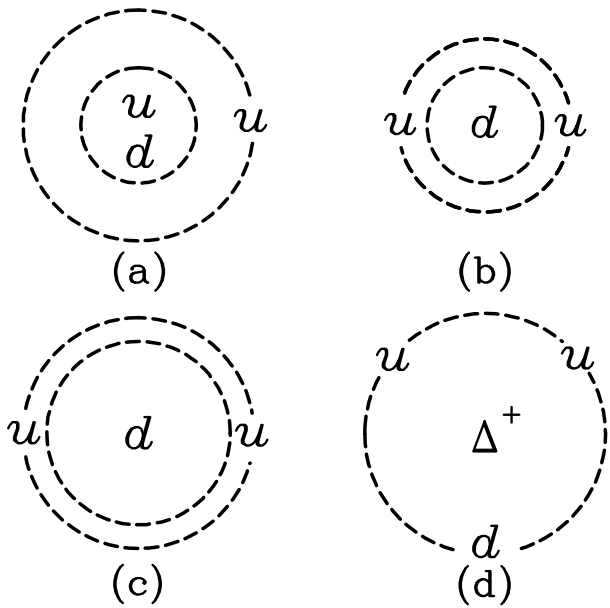

\caption{Sketches of quark distribution radii in the proton as
described by $a)$ the quark-diquark model, $b)$ the nonrelativistic
quark model and $c)$ the lattice investigations of baryon
electromagnetic form factors.  $d)$ An illustration of the quark
distributions in $\Delta^+$.  The dashed lines are representative of
the rms charge radius of the quark distributions normalized to unit
charge.
\label{proton}}
\end{figure}

\begin{figure}
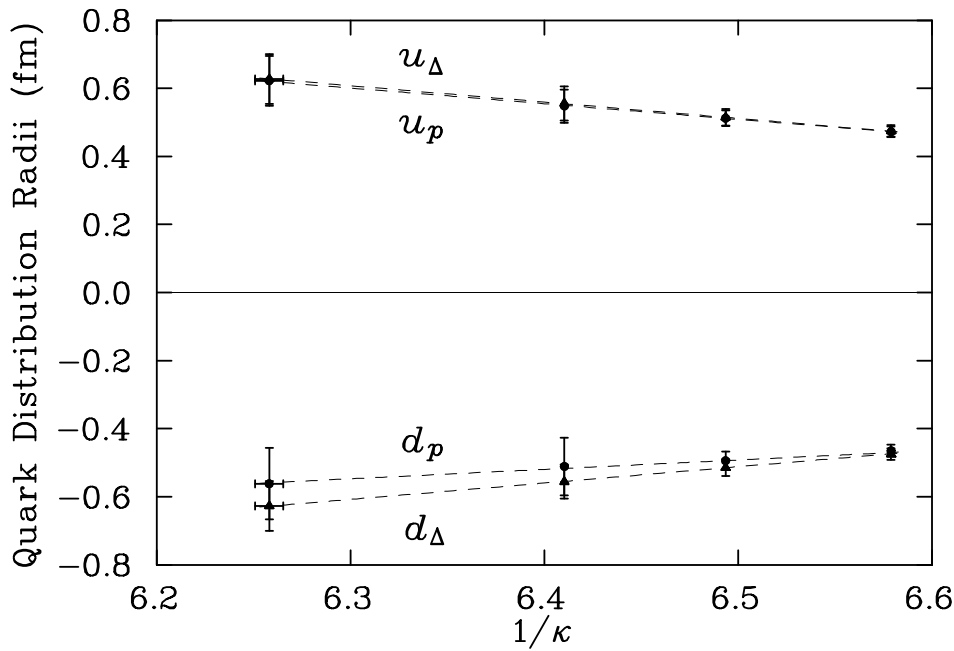

\caption{
Extrapolation of quark distribution radii in $p$ and $\Delta$
calculated at the three (rightmost) values of $\kappa$ to $\kappa_{\rm
cr}$ where $m_\pi$ vanishes.  In this and the following figure, the
lattice radii ($r/a$) have been scaled by a constant lattice spacing
of $a=0.128$ fm determined from the nucleon mass.  For clairity, the
$u$-quark radii are normalized to unit charge and $d$-quark radii are
normalized to negative unit charge.  The $u$-quark distribution radii
in $p$ and $\Delta$ are nearly identical at each value of $\kappa$.
\label{pdelrad}}
\end{figure}

\begin{figure}
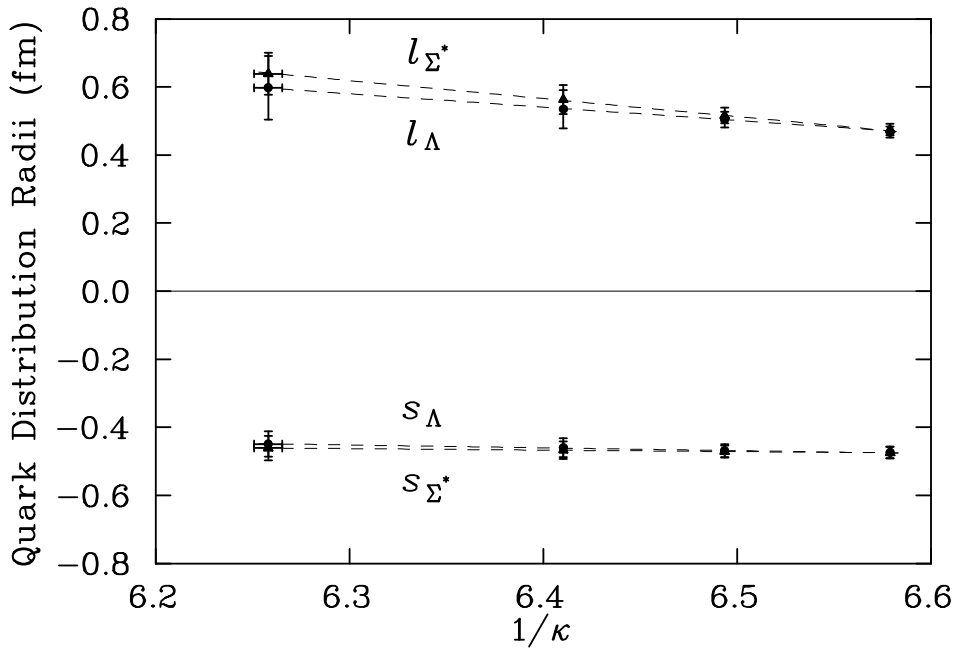

\caption{
Extrapolation of quark distribution radii in $\Lambda^0$ and
$\Sigma^{*0}$.  The distribution radius of the combined light $u$ and
$d$ quarks of $\Lambda^0$ are denoted by $l_\Lambda$ and similarly for
$\Sigma^{*0}$.  The light-quark radii are normalized to unit charge
and strange-quark radii are normalized to negative unit charge.  For
each $\kappa$, $l_\Lambda \simeq l_{\Sigma^*}$ indicating the absence
of significant scalar diquark clustering.
\label{lamsigrad}}
\end{figure}

\end{document}